# On User Association in Multi-Tier Full-Duplex Cellular Networks

Ahmed Hamdi Sakr and Ekram Hossain

## Abstract

We address the user association problem in multi-tier in-band full-duplex (FD) networks. Specifically, we consider the case of decoupled user association (DUA) in which users (UEs) are not necessarily served by the same base station (BS) for uplink (UL) and downlink (DL) transmissions. Instead, UEs can simultaneously associate to different BSs based on two independent weighted path-loss user association criteria for UL and DL. We use stochastic geometry to develop a comprehensive modeling framework for the proposed system model where BSs and UEs are spatially distributed according to independent point processes. We derive closed-form expressions for the mean rate utility in FD, half-duplex (HD) DL, and HD UL networks as well as the mean rate utility of legacy nodes with only HD capabilities in a multi-tier FD network. We formulate and solve an optimization problem that aims at maximizing the mean rate utility of the FD network by optimizing the DL and UL user association criteria. We investigate the effects of different network parameters including the spatial density of BSs and power control parameter. We also investigate the effect of imperfect self-interference cancellation (SIC) and show that it is more severe at UL, where there exist minimum required SIC capabilities for BSs and UEs for which FD networks are preferable to HD networks; otherwise, HD networks are preferable. In addition, we discuss several special cases and provide guidelines on the possible extensions of the proposed framework. We conclude that DUA outperforms coupled user association (CUA) in which UEs associate to the same BS for both UL and DL transmissions.

*Keywords*: Multi-tier cellular networks, in-band full-duplex (FD), decoupled user association (DUA), coupled user association (CUA), cell association, stochastic geometry.

## I. Introduction

In-band FD communication has recently attracted significant attention as a potential enabler for 5G networks to support higher data rates and meet the ever-increasing users demand for broadband wireless services. In contrast to HD communication in which a time-frequency resource block is only used for either transmission or reception, in-band FD communication implies simultaneous transmission and reception of information in the same frequency band [1]. This, in turn, introduces extra interference between UL and DL networks, which affects the network performance gains especially for UL transmissions that suffer from excessive DL-to-UL interference [2]. The performance of FD networks is also limited by the capability of UEs and BSs to cancel self-interference (SI), which is caused by the transmitter to its collocated receiver. Fortunately, this technology is becoming feasible thanks to the recent advancements in antenna and digital baseband technologies where SI can be reduced close to the level of noise floor in low-power devices [3].

In this paper, we focus on the user association problem in multi-tier cellular networks (i.e., networks that consist of different classes of BSs) that support in-band FD communication. We consider the general case when a cellular UE can be simultaneously served by two different BSs for UL and DL data transmission, i.e., *decoupled user association* (DUA) [4]. In addition, we aim at optimizing user association in order to maximize the mean rate utility offered by FD networks. In order to evaluate and optimize the system performance, we use a statistical approach based on stochastic geometry to capture the network randomness [5]. Specifically, in order to derive closed-form expressions for the mean rate utility in a generic link in the network, due to their analytical tractability, we use independent Poisson Point Processes (PPPs) to model the locations of BSs. The results from the analysis enable us to optimize user association to maximize the mean rate utility and to understand the impact of network parameters (such as spatial density of BSs, power control, weighting factors, and SI) on the performance to provide insightful guidelines for system design. We show that DUA is superior to *coupled user association* in which UEs associate to only one BS for both UL and DL transmissions. We also show that FD networks can outperform HD networks in terms of mean rate for sufficiently high SI cancellation (SIC) capabilities of BSs and UEs.

### A. Related Work and Motivations

In the context of DUA in multi-tier cellular networks, the authors in [6], [7] propose a framework for performance evaluation of multi-tier HD UL and DL cellular networks. In this model, the locations of BSs in each tier are modeled by independent PPPs where each network tier differs in the transmit power and spatial density. Using PPP assumption and weighted path-loss user association, [6] derives expressions for the rate coverage in HD UL and DL networks as well as the joint UL-DL rate coverage. On the other hand, closed-form expressions for the mean of the logarithm of the transmission rate and spectrum allocation for both UL and DL transmissions are derived in [7]. Furthermore, utility maximization problems are solved to optimize both user association and spectrum partitioning. The authors in [8] use stochastic geometry to derive the achievable capacity in an HD network with DUA where a real-world simulation tool (Atoll) is used to verify the accuracy of the expressions. In [9], the DUA problem is formulated as a matching game in which UEs and BSs in a two-tier FD cellular network rank one another based on some preference metric, which is a function of the achievable signal-to-interference-plus-noise ratio (SINR), in order to maximize the total throughput.

The authors in [10]–[16] evaluate the performance of FD networks using statistical modeling. In [10], a hybrid ad-hoc network is considered in which nodes with HD or FD capabilities are randomly deployed. For an ALOHA MAC protocol, it shows that FD networks achieve a $0-30\%$ higher throughput, compared to HD networks, for practical values of path-loss exponent. It also considers the effect of imperfect SIC and shows how it affects the relative performance of FD and HD transmissions. The authors in [11] present the effects of transmission duration, SIC, and ratio of HD-to-FD nodes on the performance of ad-hoc networks with asynchronous ALOHA MAC protocol. The paper also highlights different optimal operating regions in which FD networks outperform HD networks. In [12], the authors propose a system design that controls the partial overlap between UL and DL channels in order to maximize the overall rate of FD cellular networks. The paper compares the performances of two realizations for FD networks: a two-node topology (2NT) with FD UEs and BSs and a three-node topology (3NT) with FD BSs and HD UEs. It shows that 3NT achieves performance comparable to that of 2NT, which paves the road to harvest the gains of FD transmissions with HD UE terminals. In [13], a hybrid multi-tier FD cellular network is considered in which BSs operate either in FD mode or HD DL mode. PPP assumption is used to derive expressions for the successful transmission probability and network throughput. The authors in [14] consider a single-tier network under 2NT and 3NT where BSs and UEs employ directional antennas. The directional antennas are shown to manage both SI as well as co-channel interference. In [15], the authors derive the success probability and achievable spectral efficiency of FD networks under a 3NT with multi-antenna BSs and single-antenna UEs. In [16], a massive MIMO-enabled FD network is considered where BSs adopt linear zero-forcing with SI-nulling precoding while UEs adopt SI-aware power control. The authors show that massive MIMO is a viable solution for FD communications.

Although statistical modeling can be used for long-term performance evaluation of FD net-works, it does not necessarily provide sufficient insights on short-term network performance. Therefore, tools from optimization theory can be used to evaluate short-term performance of networks and to find optimal parameters that maximize certain objective functions [17]–[20]. For example, [17] proposes a joint resource management scheme to mitigate the effect of imperfect SIC in an OFDMA-based two-tier FD cellular network. This is achieved by jointly assigning UEs and transmit power for each resource block in both UL and DL transmission based on the level of SI to maximize a total utility sum of the network. The authors in [18] propose an iterative algorithm to jointly perform subcarrier assignment and power allocation in order to maximize the sum-rate performance in a single-cell FD network. In [19], the authors propose a joint UL/DL user scheduling and power allocation algorithm to maximize the system throughput by investigating the feasibility conditions of FD operation with 3NT. The authors in [20] propose two distributed power control and interference management methods to manage interference in FD networks with 3NT. The proposed MAC protocol with transmit power optimization is shown to outperform its HD counterpart in terms of total throughput.

*B. Contributions, Organization, and Notations*

In contrast to previous works on FD networks (e.g., [10]–[16]) in which CUA is used to solve the association problem, we focus on DUA as a more general association criterion for FD networks. The contributions of the paper can be summarized as follows:

- Using tools from stochastic geometry, we provide a tractable framework to analyze the performance of multi-tier FD networks with DUA. We derive closed-form expressions for the association probability, the mean interference received at UEs and BSs under weighted path-loss user association. In addition, we derive the mean rate utility of an FD network.

- We investigate the effect of FD transmissions on legacy HD terminals that do not support FD communication and we provide closed-form expressions for the mean rate utility.

- We formulate an optimization problem to maximize the mean rate utility of FD networks. By jointly optimizing both UL and DL user association to maximize the mean rate utility, we also show that DUA is superior to CUA in FD networks.

- We show how the proposed framework can be extended to different models in the literature such as traditional HD UL and HD DL networks. In addition, we highlight different tradeoffs in the network and show the effect of varying network parameters such as densities of BSs, power control parameters, and user association weighting factors on network performance.

TABLE I

List of key notations

| Notation | Definition | Notation | Definition |
|----------|-----------|----------|-----------|
| $\mathbf{\Phi}_k$ | Point process of BSs of $k$-th tier | $\lambda_k$ | Spatial density of BSs of $k$-th tier |
| $P_k$ | Transmit power of BSs of $k$-th tier | $\rho_k$ | Receiver sensitivity of BS of $k$-th tier |
| $\mathbf{\Psi}_k$ | Point process of active UEs of $k$-th tier | $\gamma_k$, $\Gamma_k$ | Transmit power of UEs of $k$-th tier |
| $P_{\max}$ | Maximum transmit power of UEs | $\epsilon$ | Power control factor of UEs |
| $U_i$ | Weighting factor for UL user association | $D_i$ | Weighting factor for DL user association |
| $\psi_{jk}$ | Joint association probability | $A_k^{\text{UL}}, A_k^{\text{DL}}$ | Per-tier association probability |
| $\tau^{\text{UL}}$, $\tau^{\text{DL}}$ | Target SINR threshold | $\sigma_{b_k}$, $\sigma_u$ | SI cancellation parameters |
| $\alpha$, $\alpha_b$, $\alpha_u$ | Path-loss exponents | $G$, $G_b$, $G_u$ | Path-loss gains |

- We investigate the effect of SI on the performance of an FD network and show that FD mode of operation is preferable to HD mode only if the SIC capabilities of UEs and BSs are sufficient to mitigate SI. Furthermore, we show that the effect of imperfect SIC is more severe at UL.

- Via Monte Carlo simulation, we validate our analytical results. Also, through numerical results we show the feasibility of FD communication to increase the mean transmission rate of cellular networks.

The rest of the paper is organized as follows. System model and assumptions are described in Section II. In Section III, both joint distance distributions and user association probabilities are derived for a typical UE. Section IV presents the analysis for FD interference as well as mean rate utility of UL, DL, and FD transmissions. An optimization problem is formulated in Section V to maximize the mean rate utility of FD transmission. Finally, numerical results and discussion are presented in Section VI before the paper is concluded in Section VII.

**Notation:** $\Gamma[s] = \int_0^\infty x^{s-1} e^{-x} \mathrm{d}x$ is the gamma function, $\Gamma[s,a] = \int_a^\infty x^{s-1} e^{-x} \mathrm{d}x$ denotes the upper incomplete gamma function, $\gamma[s,b] = \int_0^b x^{s-1} e^{-x} \mathrm{d}x$ denotes the lower incomplete gamma function, and $\Gamma[s,a,b] = \int_a^b x^{s-1} e^{-x} \mathrm{d}x$ denotes the generalized gamma function. $f(\cdot)$ and $F(\cdot)$ denote the probability density function (PDF) and cumulative distribution function (CDF), respectively. Finally, $\mathbb{E}[\cdot]$ denotes the expectation operator and $\mathbf{1}\{\cdot\}$ is the indicator function. The key mathematical notations used in this paper are summarized in Table I.

## II. System Model, Assumptions, and Methodology of Analysis

### A. Multi-Tier Network Model

We consider a cellular network that consists of $K$ tiers of BSs. We use an independent homogeneous PPP $\mathbf{\Phi}_k = \{x_{i,k} : i = 1, 2, \dots\}$ with spatial density $\lambda_k$ BS/m$^2$ to model locations of BSs belonging to the $k$-th tier where $1 \le k \le K$ and $x_{i,k} \in \mathbb{R}^2$ denotes the location of the $i$-th BS in that tier. For all BSs in $k$-th tier, transmit power is fixed and equal to $P_k$. Locations of UEs are modeled in $\mathbb{R}^2$ according to an arbitrary independent point process $\mathbf{\Phi}_U$ with a spatial density $\lambda_u >> \sum_{k=1}^{K} \lambda_k$. Saturation condition is assumed where each transmitter (i.e., a BS or a UE) sends data packets at the beginning of each time slot in a time-slotted transmission scenario.

### B. Channel Model

We assume a co-channel deployment where the wireless channels are subject to both large-scale (i.e., path-loss) and small-scale fading. Let $\|x - y\|$ be the propagation distance between a generic transmitter (i.e., a BS or a UE) located at $x$ and a receiver located at $y$. The path-loss of this link is defined as $L(x,y) = \bar{G}\|x - y\|^{\bar{\alpha}}$, where $\bar{\alpha} > 2$ is the path-loss exponent and $\bar{G} > 0$ is a constant gain. In this work, we assume different path-loss exponents and gains for different communication links [21]. That is, $\bar{\alpha} = \alpha$ and $\bar{G} = G$ for BS-UE links, $\bar{\alpha} = \alpha_b$ and $\bar{G} = G_b$ for BS-BS links, and $\bar{\alpha} = \alpha_u$ and $\bar{G} = G_u$ for UE-UE links.

In addition to the distance-dependent path-loss, the small-scale fading component of a channel is modeled by Rayleigh fading with unit average power where different links are assumed to be independent and identically distributed (i.i.d.). Hence, the power gain of all channels is exponentially-distributed and denoted by $h \sim \text{Exp}(1)$[1]. Channel coherence time is greater than or equal to the frame duration. SI channel is modeled by Nakagami-$m$ fading with parameters $(m_{b_k}, \sigma_{b_k})$ and $(m_u, \sigma_u)$ for BSs from $k$-th tier and UEs, respectively [16], [22]. Hence, the power gain of SI channel follows a Gamma distribution and denoted by $h_b \sim \text{Gamma}(m_{b_k}, \frac{\sigma_{b_k}}{m_{b_k}})$ for BSs and $h_u \sim \text{Gamma}(m_u, \frac{\sigma_u}{m_u})$ for UEs, where $\frac{1}{\sigma_{b_k}}$ and $\frac{1}{\sigma_u}$ are SIC capabilities of BSs and UEs, respectively. There is no intra-cell interference between UL (DL) transmissions where different UEs in a cell are served in UL (DL) using orthogonal time-frequency resources (e.g. OFDMA). Hence, there are only one active UE in UL and one active UE in DL per BS at a certain time slot and channel.

### C. Power Control Model

The UEs adopt fractional channel inversion power control where a UE at $x$ adjusts its transmit power to $\rho_k(G\|x - y\|^{\alpha})^{\epsilon}$ to compensate for the large-scale fading such that the average received

---

[1]This assumption can be relaxed to other scenarios, e.g., Gamma-distributed power envelope assumptions to model multi-antenna transmissions.

| Mode | Definition |
|------|------------|
| FD | All BSs and UEs are FD |
| 3NT FD | All BSs are FD and all UEs are HD |
| Legacy DL | A typical HD UE with DL transmissions in an FD network |
| Legacy UL | A network tier of HD BSs with UL transmissions in an FD network |
| HD DL | All BSs and UEs are HD with DL transmissions |
| HD UL | All BSs and UEs are HD with UL transmissions |

signal power at the serving BS at $y$ is equal to $\rho_k(G\|x - y\|^{\alpha})^{-(1-\epsilon)}$. Note that $0 \le \epsilon \le 1$ is the power control factor and $\rho_k$ is the open loop power spectral density (or receiver sensitivity). Here, we use $\gamma_k$ to denote the instantaneous transmit power of a UE transmitting to a BS belonging to the $k$-th tier where $\Gamma_k$ is the corresponding random variable. UEs have a limited transmit power budget of $P_{\max}$, where the UEs that are unable to perform channel inversion, transmit with maximum power. Other power control mechanisms such as SI-aware power control [16] and interference-aware power control [23] can also be adopted.

### D. Mode of Operation and User Association

Besides FD links in which a UE simultaneously transmits and receives data in the same channel, we consider network tiers in which UEs or BSs do not support FD transmissions (i.e., HD UEs and HD BSs). That is, we also evaluate the performance of the network when (i) there is a typical HD UE that can *only receive DL transmission* from one BS in a channel during a transmission interval, and (ii) there is a network tier of HD BSs each of which can *only receive UL transmission* in a channel from one UE during a transmission interval. A communication link is referred to as a *legacy uplink* when the UE-BS channel carries data only from the UE to the serving HD BS. On the other hand, when the UE-BS channel carries data only from the BS to the HD UE being served, the link is referred to as a *legacy downlink*. Furthermore, we consider the 3NT model, where the network consists of FD BSs each of which serves two HD UEs: one HD UE in DL and one HD UE in UL. The aforementioned scenarios are summarized in Table II and will be discussed in details in Sections IV and VI.

For FD UEs, we consider the case where user association in UL is decoupled from that in DL. Hence, an FD UE is *not necessarily* served by the same BS in both UL and DL. That is, a UE may simultaneously receive data from one BS and transmit data to another in the same channel. To illustrate, without loss of generality, Fig. 1 shows a realization of a two-tier FD cellular network where a macro-cell network tier is overlaid with a denser and lower power small-cell network tier. It shows that the coverage area of each cell in DL is different from that in UL. For example, although UE 1 is served by a macro-cell BS in DL, it is located in UL coverage area of another small-cell BS, hence, served by two different BSs. On the other hand, UE 2 is located in the coverage area of one small-cell BS in both DL and UL, hence, served by the same BS. Fig. 1 also shows an example of 3NT realization where BS 3 serves two HD UEs simultaneously: one HD UE is in DL and another HD UE is in UL.

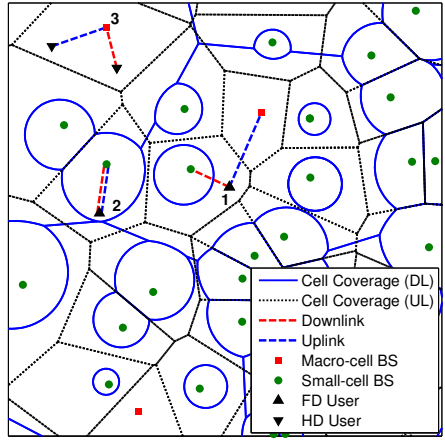

Fig. 1. A two-tier FD cellular network with macro-cell and small-cell BSs. Solid lines show DL coverage area while dotted lines show UL coverage of each cell.

We assume a weighted path-loss user association criterion similar to [24], [25] where each UE independently associates with the BS(s) that minimizes the weighted path-loss of UL and/or DL. That is, for a UE at $y$, with a slight abuse of notation, let $x_i$, $x^{\mathrm{UL}}$, and $x^{\mathrm{DL}}$ denote the BS with minimum path-loss from $i$-th tier, serving BS in UL, and serving BS in DL, respectively. Then, the association criterion for UL and DL can be described, respectively, as follows:

$$x^{\mathrm{UL}} = \arg\min_{x \in \{x_i\}} U_i \|x - y\|^\alpha \tag{1}$$

$$x^{\mathrm{DL}} = \arg\min_{x \in \{x_i\}} D_i \|x - y\|^\alpha \tag{2}$$

where $x_i = \arg\min_{x \in \Phi_i} \|x - y\|^\alpha$, $i = \{1, 2, \ldots, K\}$, and $U_i$ and $D_i$ are the weighting factors, respectively, for UL and DL user association for the $i$-th tier.

*Assumption* 1. Let $\mu_i = \frac{U_i}{D_i}$ and without loss of generality, let us assume that the network tiers are ordered such that $\mu_1 \leq \mu_2 \leq \cdots \leq \mu_K$.

Note that $U_i$ and $D_i$ are design parameters and are not necessarily equal. Varying thsese weighting factors can result in different association scenarios. For example, (i) *coupled user association*: when $U_i = D_i$, each UE associates to the same BS for both DL and UL, (ii) *minimum-distance user association*: when $U_i = U$ (or $D_i = D$), each UE associates to the nearest BS for UL (or DL), (iii) *maximum-received power user association*: when $D_i = P_i^{-1}$, each UE associates to the BS that offers the strongest received power for DL, and (iv) *minimum-transmit power user association*: when $U_i = \rho_i$, each UE associates to the BS that requires lowest transmit power for UL.

### E. Methodology of Analysis

Based on the system model described above, we aim at quantifying the performance of a generic FD link in terms of mean rate in nats/sec/Hz. We first derive the user association probability and distance distributions based on decoupled and weighted path-loss user association. Then, we obtain the mean of interference experienced at a typical BS in UL and at a typical UE in DL. Next, we derive the mean rate utility of the network as well as that of UL and DL transmissions. Finally, we optimize UL and DL weighting factors such that the mean rate utility of the FD network is maximized.

### III. Analyses of Distance and User Association Probabilities in FD Networks

### A. Analysis of User Association Probabilities

Note that, even with DUA, an FD UE can associate to the same BS in the $k$-th tier for both UL and DL transmissions with a certain probability. It is worth mentioning that this scenario is different from the CUA as it does not necessitate $U_k$ and $D_k$ to be equal (as in the CUA). That is, although the UE still uses two different decision criteria for UL and DL user association (i.e., DUA in (1) and (2)), one BS meets both criteria. This event occurs depending on the network realization. Let $\psi_{jk}$ denote the *joint association probability* that a UE is served by a BS from the $j$-th tier in DL and a BS from the $k$-th tier in UL. The following lemma characterizes this probability.

**Lemma 1.** *(Joint association probability)* The probability that an FD UE with DUA is served by the $j$-th tier for DL and $k$-th tier for UL transmissions is

$$\psi_{jk} = \begin{cases} \lambda_j \left( \sum\limits_{i=1}^{K} \max\left\{ \mathcal{D}_{ji}, \mathcal{U}_{ji} \right\}^{\frac{2}{\alpha}} \lambda_i \right)^{-1}, & j = k \\ \lambda_j \lambda_k \left( \dfrac{D_j}{U_k} \right)^{\frac{2}{\alpha}} \sum\limits_{l=k}^{j-1} \dfrac{1}{\Upsilon_l^2(j)} \left( \dfrac{\mu_{l+1}^{\frac{2}{\alpha}}}{1+\mu_{l+1}^{\frac{2}{\alpha}}\Omega_l} - \dfrac{\mu_l^{\frac{2}{\alpha}}}{1+\mu_l^{\frac{2}{\alpha}}\Omega_l} \right), & k < j \\ 0, & k > j \end{cases} \tag{3}$$

where $\mathcal{U}_{jk} = \frac{U_j}{U_k}$, $\mathcal{D}_{jk} = \frac{D_j}{D_k}$, $\Upsilon_l(j) = \sum\limits_{i=l+1}^{K} \mathcal{D}_{ji}^{\frac{2}{\alpha}} \lambda_i$, and $\Omega_l = \left( \sum\limits_{i=1}^{l} \frac{\lambda_i}{U_i^{\frac{2}{\alpha}}} \right) \left( \sum\limits_{i=l+1}^{K} \frac{\lambda_i}{D_i^{\frac{2}{\alpha}}} \right)^{-1}$.

*Proof:* See **Appendix A-I**. ∎

Note that the joint association probability $\psi_{jk}$ is different from the *per-tier association probability*, which is defined as the probability that a UE is served by a BS belonging to a certain tier for DL (UL) regardless of the serving UL (DL) BS. This per-tier association probability can be obtained directly from the joint association probability derived in **Lemma 1**. The following lemma provides expressions for this probability for both DL and UL transmissions.

**Lemma 2.** *(Per-tier association probability)* The probability that a UE associates to a BS from the $j$-th tier for either DL or UL is defined, respectively, as follows:

$$A_j^{\mathrm{DL}} = \lambda_j (\Lambda_j^{\mathrm{DL}})^{-1} \quad \text{and} \quad A_j^{\mathrm{UL}} = \lambda_j (\Lambda_j^{\mathrm{UL}})^{-1} \tag{4}$$

where $\Lambda_j^{\mathrm{DL}} = \sum_{i=1}^{K} \mathcal{D}_{ji}^{\frac{2}{\alpha}} \lambda_i$ and $\Lambda_j^{\mathrm{UL}} = \sum_{i=1}^{K} \mathcal{U}_{ji}^{\frac{2}{\alpha}} \lambda_i$ are the *effective spatial densities* of the $j$-th tier for DL and UL transmissions, respectively.

## B. Analysis of Distance to Serving BS(s)

Based on the system model and user association criteria described above, for UEs operating in FD mode, the marginal PDFs of the distances to the serving BSs in DL and UL are presented in the following lemma the proof of which can be found in [25, Appendix A].

**Lemma 3.** *(Marginal distance distributions)* The CDF of the distance between a generic UE associated with the $j$-th tier for DL (or UL) and its serving BS is $\mathbb{P}[R_j^m \leq r] = 1 - \exp\left[ -\pi \Lambda_j^m r^2 \right]$ and the $n$-th moment is $\mathbb{E}_{R_j^m}\left[ R_j^n \right] = \Gamma\left[ \frac{2+n}{2} \right] \left( \pi \Lambda_j^m \right)^{-\frac{n}{2}}$, where $m \in \{\mathrm{DL}, \mathrm{UL}\}$.

Furthermore, the joint PDF of the distance to serving BSs for DL and UL transmissions for FD UEs is presented in the following lemma.

**Lemma 4.** *(Joint distance distribution)* The joint PDF of the distance(s) from a generic FD UE to its serving BS(s), when associated to the $j$-th tier for DL and $k$-th tier for UL, is

$$f_{\mathbf{R}}(r_j, r_k) = \begin{cases} 2\pi\lambda_j r_j \exp\left[-\pi\sum_{i=1}^{K}\max\left\{\mathcal{D}_{ji},\mathcal{U}_{ji}\right\}^{\frac{2}{\alpha}}\lambda_i r_j^2\right], & j=k, r_k=r_j \\ 4\pi^2\lambda_j\lambda_k r_j r_k \exp\left[-\pi\sum_{i=1}^{K}\max\left\{\mathcal{D}_{ji}r_j^\alpha,\mathcal{U}_{ki}r_k^\alpha\right\}^{\frac{2}{\alpha}}\lambda_i\right], & k<j, (r_j,r_k)\in\mathcal{A} \end{cases} \tag{5}$$

where $\mathcal{A} = \left\{(r_j, r_k) : r_j \geq 0, \mathcal{D}_{jk}^{\frac{1}{\alpha}} r_j < r_k < \mathcal{U}_{jk}^{\frac{1}{\alpha}} r_j\right\}$, and their expectations are given by

$$\begin{aligned}
\mathbb{E}_{\mathbf{R}}[R_j^m] &= \Gamma\left[\tfrac{2+m}{2}\right]\left(\pi\sum_{i=1}^{K}\max\left\{\mathcal{D}_{ji},\mathcal{U}_{ji}\right\}^{\frac{2}{\alpha}}\lambda_i\right)^{-\frac{m}{2}}, & j=k \\
\mathbb{E}_{\mathbf{R}}[R_j^m] &= \left(\sum_{l=k}^{j-1}\mathcal{H}_{jl}\left[\mu_l, \mu_{l+1}; 0\right]\right)^{-1}\sum_{l=k}^{j-1}\tfrac{-1}{\Omega_l}\mathcal{H}_{jl}\left[1, 1; m\right], & k<j \\
\mathbb{E}_{\mathbf{R}}[R_k^n] &= \left(\sum_{l=k}^{j-1}\mathcal{H}_{jl}\left[\mu_l, \mu_{l+1}; 0\right]\right)^{-1}\sum_{l=k}^{j-1}\mathcal{H}_{jl}\left[\mu_l, \mu_{l+1}; n\right], & k<j
\end{aligned} \tag{6}$$

where $\mathcal{H}_{jl}\left[a, b; c\right] = \frac{\Gamma\left[\frac{2+c}{2}\right]}{\pi^{\frac{c}{2}}}\left(\Upsilon_l(j)\right)^{-\frac{4+c}{2}}\left[a^{\frac{2+c}{\alpha}}\left(1+\mu_l^{\frac{2}{\alpha}}\Omega_l\right)^{-\frac{2+c}{2}}-b^{\frac{2+c}{\alpha}}\left(1+\mu_{l+1}^{\frac{2}{\alpha}}\Omega_l\right)^{-\frac{2+c}{2}}\right]$.

*Proof:* See **Appendix A-II**. ∎

## C. Analysis of Uplink Transmission Power

As stated above, the UEs served in UL by a BS from $k$-th tier are assumed to perform fractional channel inversion with open loop power spectral density $\rho_k$. The UEs are also assumed to have a constraint $P_{\max}$ on the transmit power. Thus, we define the required amount of transmit power of a UE when it associates with a BS from the $k$-th tier as $\gamma_k = \min\left\{\rho_k G^\epsilon R_k^{\epsilon\alpha}, P_{\max}\right\}$ where $R_k$ is the distance to the serving BS from tier $k$ for UL transmission and its CDF is given in **Lemma 3**. Therefore, following **Lemma 3**, we derive the CDF of the transmit power where the proof follows directly such that $\mathbb{P}[\Gamma_k > t] = \mathbb{P}\left[R_k^{\text{UL}} > \left(\frac{t}{\rho_k G^\epsilon}\right)^{\frac{1}{\epsilon\alpha}}\right]\cdot\mathbf{1}\{t < P_{\max}\}$.

**Lemma 5.** *(Transmit power distribution)* The CDF of the transmit power of a UE associated to tier $k$ in UL and performing fractional channel inversion power control is $\mathbb{P}[\Gamma_k \leq t] = 1 - \exp\left[-\pi\Lambda_k^{\text{UL}}\left(\frac{t}{\rho_k G^\epsilon}\right)^{\frac{2}{\epsilon\alpha}}\right]\cdot\mathbf{1}\{t < P_{\max}\}$ and its $n$-th moment is given by

$$\mathbb{E}[\Gamma_k^n] = \frac{n\epsilon\alpha\rho_k^n G^{n\epsilon}}{2(\pi\Lambda_k^{\text{UL}})^{\frac{n\epsilon\alpha}{2}}}\gamma\left[\frac{n\epsilon\alpha}{2}, \pi\Lambda_k^{\text{UL}}\left(\frac{P_{\max}}{\rho_k G^\epsilon}\right)^{\frac{2}{\epsilon\alpha}}\right]. \tag{7}$$

## IV. Analysis of Rate Coverage of FD Transmissions

In this section, we characterize the mean rate utility for a generic FD link. We assume that during a transmission interval, a receiver (i.e., BS or a UE) using a particular channel to communicate with its corresponding transmitter experiences interference from all other BSs as well as UEs reusing the same channel. We start by defining the SINR received at a typical UE and BS. Then, the mean rate utility is also derived for a generic FD link and several special cases in the literature.

### A. Signal-to-Interference-plus-Noise Ratio

For the system model described above, the SINR can be expressed as follows:

$$\text{SINR}_{jk}^{\text{DL}} = \frac{P_j h}{(\text{I}^{\text{DL}} + \text{I}_{\text{SI}}^{\text{DL}} + \sigma^2) G \|x^{\text{DL}}\|^{\alpha}} \quad \text{and} \quad \text{SINR}_{jk}^{\text{UL}} = \frac{\min\{\rho_k G^{\epsilon} \|x^{\text{UL}}\|^{\epsilon\alpha}, P_{\max}\} h}{(\text{I}^{\text{UL}} + \text{I}_{\text{SI}}^{\text{UL}} + \sigma^2) G \|x^{\text{UL}}\|^{\alpha}} \quad (8)$$

where $x^{\text{DL}} \in \Phi_j$ and $x^{\text{UL}} \in \Phi_k$ denote the serving BS for DL and UL transmissions, respectively, as defined in (1)-(2). $\sigma^2$ is the additive noise power, $\{h\}$ are channel power gains where subscripts are dropped for simplicity, and $\text{I}^{\text{DL}}$ and $\text{I}^{\text{UL}}$ are the interference received, respectively, at the tagged UE (i.e., DL) and BS (i.e., UL) from all other BSs and UEs sharing the same channel. Note that, unlike HD networks, interference signals received from DL and UL transmissions are correlated due to the relative distance between each UE-BS pair and the observation point. In order to characterize the interference, using the fact that each cell has exactly one DL and one UL transmissions, we pair each UE and its serving BS in UL together as an interfering pair. Hence, the aggregate interference at a typical receiver (i.e., a UE or a BS) located at the origin $(0,0)$ can be defined as follows[2]:

$$\text{I}^{\text{DL}} = \sum_{i=1}^{K} \sum_{x \in \Phi_i \setminus \{x^{\text{DL}}\}} \frac{P_i h}{G \|x\|^{\alpha}} + \sum_{y \in \Psi_i \setminus \{u[x^{\text{UL}}]\}} \frac{\gamma_i(y) h}{G_u \|y\|^{\alpha_u}} \quad (9)$$

$$\text{I}^{\text{UL}} = \sum_{i=1}^{K} \sum_{x \in \Phi_i \setminus \{x^{\text{UL}}\}} \frac{P_i h}{G_b \|x\|^{\alpha_b}} + \sum_{y \in \Psi_i \setminus \{u[x^{\text{UL}}]\}} \frac{\gamma_i(y) h}{G \|y\|^{\alpha}} \quad (10)$$

where $u[x]$ returns the UE served by BS $x$ in UL, $\gamma_i(y) = \min\{\rho_i G^{\epsilon} \|y - u^{-1}[y]\|^{\epsilon\alpha}, P_{\max}\}$ is the transmit power of the UE, and $\Psi_i$ is the point process that represents the active UEs. Note that, due to the correlation between the positions of active UEs and BSs resulting from the orthogonal time-frequency resource allocations (e.g., OFDMA), this point process does is not

---

[2]Due to the homogeneous PPP assumption, interference statistics are independent of the observation point [5].

a PPP. Instead, the positions of active UEs can be seen as a Voronoi perturbed lattice process which is not mathematically tractable [6], [26].

*Assumption 2.* The active UEs from the $i$-th tier are assumed to form an arbitrary point process $\Psi_i$ that is (i) stationary with spatial density $\lambda_i$, (ii) independent of the point processes of active UEs from different tiers, and (iii) independent of $\Phi_i{}^3$.

### B. Interference Model

For each type of tagged receivers (i.e., BSs or UEs) and interferers (i.e., BSs or UEs), we define a pair correlation function $g(r)$ which in turn quantifies the spatial density of interfering nodes from $i$-th tier as $\lambda_i g(r)$.

(i) **BSs-to-UEs (DL-to-DL):** Using (2), we know that no interfering BS from $i$-th tier can have less weighted path-loss to a UE than its serving BS, hence, $\|x\| > \mathcal{D}_{ji}^{\frac{1}{\alpha}}\|x^{\mathrm{DL}}\|$, where $x \in \boldsymbol{\Phi}_i$ and $x^{\mathrm{DL}} \in \boldsymbol{\Phi}_j$ [5]–[7], [24], [27]. Therefore, $g_1(r) = \mathbf{1}\{r > \mathcal{D}_{ji}^{\frac{1}{\alpha}} R_j\}$, where $R_j = \|x^{\mathrm{DL}}\|$.

(ii) **UEs-to-BS (UL-to-UL):** There is no exact boundary for the exclusion region around the tagged BS where interfering UEs can be arbitrarily close. However, using (1), we know that none of the interfering UEs associates with that BS, hence, for a UE $u$ served by a BS from $i$-th tier, $\|x^{\mathrm{UL}} - u\| > \mathcal{U}_{ik}^{\frac{1}{\alpha}}\|x - u\|$, where $x \in \boldsymbol{\Phi}_i$ and $x^{\mathrm{UL}} \in \boldsymbol{\Phi}_k$ [6], [7], [25], [28]. Therefore, $g_2(r) = \mathbf{1}\{r > \mathcal{U}_{ik}^{\frac{1}{\alpha}} R_i\}$, where $R_i = \|x - u\|$.

(iii) **BSs-to-BS (DL-to-UL):** Due to the PPP assumption, the BSs can be very close to each other, however, this is not true in practice due to physical constraints, different antenna heights, etc. [21]. Here, we use the approximation proposed in [29] to model interfering BSs from $i$-th tier as a non-homogeneous point process with $g_3(r) = (1 - \exp[-\pi\frac{\lambda_i}{\beta_b}r^2])\mathbf{1}\{r > d_b\}$ to model the repulsion between BSs in practical deployments where $\beta_b$ is the repulsion parameter and $d_b$ is a constraint on the length of an DL-to-UL interference link.

(iv) **UEs-to-UE (UL-to-DL):** We use a similar approximation to model the interfering UEs served by the $i$-th tier of BSs in UL as a non-homogeneous point process with $g_4(r) = (1 - \exp[-\pi\frac{\lambda_i}{A_i^{\mathrm{UL}}}r^2])\mathbf{1}\{r > d_u\}$, where $A_i^{\mathrm{UL}}$ is the repulsion parameter as in [6] and $d_u$ is the minimum length of an UL-to-DL interference link [30].

---

³Other approximations for $\Psi_i$ exist in the literature, e.g., non-homogeneous PPP with spatial density $\lambda_i\left(1 - \exp[-\pi\Lambda_i^{\mathrm{UL}}r^2]\right)$ [6], or more generally $\lambda_i g(r)$, where $g(r)$ is a pair correlation function as proposed in [29]. Another approximation is using PPP assumption with exclusion ball of radius $\sqrt{\frac{1}{\pi\Lambda_i^{\mathrm{UL}}}}$ such as in [31] for single-tier networks.

## C. Self-Interference

In (8), $\mathsf{I}_{\mathrm{SI}}^{\mathrm{DL}}$ is SI resulting from UL transmission at the UE, and $\mathsf{I}_{\mathrm{SI}}^{\mathrm{UL}}$ is SI resulting from DL transmission at the BS. Since the SI incurred at a given receiver depends on its own transmit power, we define the residual SI power after cancellation as follows:

$$\mathsf{I}_{\mathrm{SI}}^{\mathrm{DL}} = \min\left\{\rho_k G^\epsilon \|x^{\mathrm{UL}}\|^{\epsilon\alpha}, P_{\max}\right\} h_u \quad \text{and} \quad \mathsf{I}_{\mathrm{SI}}^{\mathrm{UL}} = P_k h_{b_k} \tag{11}$$

where $h_{b_k}$ and $h_u$ represent SI channels such that $\sigma_{b_k} = \mathbb{E}[h_{b_k}]$ and $\sigma_u = \mathbb{E}[h_u]$ are the inverse of SIC capability of BSs from tier $k$ and UEs, respectively.

## D. Mean Rate Utility

For a given FD link, let us define the rate coverage probability as the probability of UL (DL) transmission rate to be higher than a required target rate threshold $\mathsf{R}_o^{\mathrm{UL}}$ ($\mathsf{R}_o^{\mathrm{DL}}$). In other words, for UL (DL) transmission to achieve the target rate, the SINR received at the BS (UE) must be greater than a prescribed threshold $\tau^{\mathrm{UL}}$ ($\tau^{\mathrm{DL}}$) that can be given using Shannon's formula such that $\tau^m = \exp[\mathsf{R}_o^m] - 1, m \in \{\mathrm{DL}, \mathrm{UL}\}$. Hence, the rate $\mathsf{R}_{jk}$ of a generic FD link measured in nats/sec/Hz for a UE served by a BS from $j$-th tier for DL and a BS from $k$-th tier for UL is defined as:

$$\mathsf{R}_{jk} = \underbrace{\mathbb{P}(\mathsf{SINR}_{jk}^{\mathrm{DL}} > \tau^{\mathrm{DL}})\ln\left[1 + \tau^{\mathrm{DL}}\right]}_{=\mathsf{R}_{jk}^{\mathrm{DL}}} + \underbrace{\mathbb{P}(\mathsf{SINR}_{jk}^{\mathrm{UL}} > \tau^{\mathrm{UL}})\ln\left[1 + \tau^{\mathrm{UL}}\right]}_{=\mathsf{R}_{jk}^{\mathrm{UL}}} \tag{12}$$

which corresponds to a fixed data rate transmission (i.e., $\mathsf{R}_o^{\mathrm{DL}} = \ln[1 + \tau^{\mathrm{DL}}]$ and $\mathsf{R}_o^{\mathrm{UL}} = \ln[1 + \tau^{\mathrm{UL}}]$) when the SINRs for DL and UL transmissions (i.e., $\mathsf{SINR}_{jk}^{\mathrm{DL}}$ and $\mathsf{SINR}_{jk}^{\mathrm{UL}}$, respectively) exceed the predefined thresholds; otherwise, the transmission rate is zero.

Furthermore, we define the mean rate utility of a generic FD link in the network as:

$$\bar{\mathsf{R}} = \sum_{j=1}^{K}\sum_{k=1}^{K} \psi_{jk} f\left(\mathsf{R}_{jk}^{\mathrm{DL}}, \mathsf{R}_{jk}^{\mathrm{UL}}\right) \tag{13}$$

where $f\left(\mathsf{R}_{jk}^{\mathrm{DL}}, \mathsf{R}_{jk}^{\mathrm{UL}}\right)$ is the mean utility of an FD link given that the UE is associated with $j$-th tier for DL and $k$-th tier for UL. In this work, we use $f\left(\mathsf{R}_{jk}^{\mathrm{DL}}, \mathsf{R}_{jk}^{\mathrm{UL}}\right) = \mathbb{E}\left[\ln\left[\mathsf{R}_{jk}^{\mathrm{DL}}\mathsf{R}_{jk}^{\mathrm{UL}}\right]\right] = \mathbb{E}\left[\ln\mathsf{R}_{jk}^{\mathrm{DL}}\right] + \mathbb{E}\left[\ln\mathsf{R}_{jk}^{\mathrm{UL}}\right]$ as the mean utility function measured in ln[nats/sec/Hz] [7], [32]. The motivation of considering this utility is to achieve *proportional fairness* among UEs. This is achieved by improving the rates of links with low rate coverage probability while saturating the rates of links with high rate coverage probability. This performance metric is widely used in the literature to solve the problem of radio resource allocations while achieving fairness via maximizing $\bar{R}$ [33]. Therefore, hereafter, we derive the following performance metric for a generic UE operating in the FD mode:

$$\mathbb{E}[\ln R_{jk}^m] = \ln R_o^m + \mathbb{E}\left[\ln\left[\mathbb{P}(\mathsf{SINR}_{jk}^m > \tau^m)\right]\right], \qquad m \in \{\mathsf{DL}, \mathsf{UL}\} \tag{14}$$

where the expectation is with respect to $h$, $\mathbf{R}$, $\mathbf{\Phi}_i$, and $\mathbf{\Psi}_i$.

First, we focus on UL rate coverage, which can be obtained as follows:

$$\mathbb{E}\left[\ln\left[\mathbb{P}(\mathsf{SINR}_{jk}^{\mathsf{UL}} > \tau^{\mathsf{UL}})\right]\right] \overset{(a)}{=} \mathbb{E}\left[\ln\left[\mathbb{P}\left(g > \frac{\tau^{\mathsf{UL}}G\left(\mathsf{I}^{\mathsf{UL}} + \mathsf{I}_{\mathsf{SI}}^{\mathsf{UL}} + \sigma^2\right)}{\min\{\rho_k G^\epsilon R_k^{\epsilon\alpha}, P_{\max}\}R_k^{-\alpha}}\right)\right]\right]$$

$$\overset{(b)}{=} -\tau^{\mathsf{UL}}G \ \mathbb{E}_{\mathbf{R}}\left[\frac{\mathbb{E}\left[\mathsf{I}^{\mathsf{UL}}\right] + \mathbb{E}\left[\mathsf{I}_{\mathsf{SI}}^{\mathsf{UL}}\right] + \sigma^2}{\min\{\rho_k G^\epsilon R_k^{\epsilon\alpha}, P_{\max}\}R_k^{-\alpha}}\right] \tag{15}$$

where $(a)$ follows from (8) and $(b)$ follows from the Rayleigh fading assumption. The mean of $\mathsf{I}^{\mathsf{UL}}$ and $\mathsf{I}_{\mathsf{SI}}^{\mathsf{UL}}$ are presented in the following lemma.

***Lemma 6.*** *(Mean of uplink interference)* The mean interference power received at a typical BS belonging to the $k$-th tier is given by

$$\mathbb{E}\left[\mathsf{I}^{\mathsf{UL}}\right] = \sum_{i=1}^{K} 2\pi\lambda_i \left(\frac{\mathcal{K}_1(d_b, \alpha_b, \frac{\lambda_i}{\beta_b})}{G_b}P_i + \frac{\mathcal{U}_{ik}^{\frac{2-\alpha}{\alpha}}\mathcal{K}_2(i)}{G(\alpha-2)}\rho_i\right) \tag{16}$$

and the mean of SI is $\mathbb{E}\left[\mathsf{I}_{\mathsf{SI}}^{\mathsf{UL}}\right] = \sigma_{b_k}P_k$, where

$$\mathcal{K}_1(d, \alpha, \Lambda) = \frac{d^{2-\alpha}}{\alpha-2} - \frac{1}{2}\left(\pi\Lambda\right)^{\frac{\alpha-2}{2}}\Gamma\left[\frac{2-\alpha}{2}, \pi\Lambda d^2\right] \overset{\substack{(2<\alpha<4)\\(d\to 0)}}{=} -\frac{1}{2}\left(\pi\Lambda\right)^{\frac{\alpha-2}{2}}\Gamma\left[\frac{2-\alpha}{2}\right] \tag{17}$$

and

$$\mathcal{K}_2(i) = (\pi\Lambda_i^{\mathsf{UL}})^{\frac{\alpha(1-\epsilon)-2}{2}}\Gamma\left[\frac{4-\alpha(1-\epsilon)}{2}, \pi\Lambda_i^{\mathsf{UL}}d_o^2, \pi\Lambda_i^{\mathsf{UL}}\left(\frac{P_{\max}}{\rho_i G^\epsilon}\right)^{\frac{2}{\epsilon\alpha}}\right]$$

$$+ (\pi\Lambda_i^{\mathsf{UL}})^{\frac{\alpha-2}{2}}\frac{P_{\max}}{\rho_i G^\epsilon}\Gamma\left[\frac{4-\alpha}{2}, \pi\Lambda_i^{\mathsf{UL}}\left(\frac{P_{\max}}{\rho_i G^\epsilon}\right)^{\frac{2}{\epsilon\alpha}}\right]$$

$$\overset{(P_{\max}\to\infty)}{=} (\pi\Lambda_i^{\mathsf{UL}})^{\frac{\alpha(1-\epsilon)-2}{2}}\Gamma\left[\frac{4-\alpha(1-\epsilon)}{2}, \pi\Lambda_i^{\mathsf{UL}}d_o^2\right]. \tag{18}$$

in which $d_o$ is the minimum distance between UEs and BSs [21].

*Proof:* See **Appendix B-I**. ∎

Note that in (18), the first term represents interference from UEs that are able to perform channel inversion power control without exceeding the power budget where the second term represents the interference from UEs who are transmitting with the maximum power $P_{\max}$.

Next, similar to UL, the rate coverage for DL can be obtained as follows:

$$\mathbb{E}\left[\ln\left[\mathbb{P}(\mathsf{SINR}_{jk}^{\mathrm{DL}} > \tau^{\mathrm{DL}})\right]\right] = -\tau^{\mathrm{DL}}G\,\mathbb{E}_{\mathbf{R}}\left[\frac{\mathbb{E}\left[\mathsf{I}^{\mathrm{DL}}\right] + \mathbb{E}\left[\mathsf{I}_{\mathrm{SI}}^{\mathrm{DL}}\right] + \sigma^2}{P_j R_j^{-\alpha}}\right] \tag{19}$$

where the mean of $\mathsf{I}^{\mathrm{DL}}$ and $\mathsf{I}_{\mathrm{SI}}^{\mathrm{DL}}$ are presented in the following lemma.

**Lemma 7.** *(Mean of downlink interference)* The mean interference power received at a typical UE served in DL by the $j$-th tier and in UL by the $k$-th tier is given by

$$\mathbb{E}\left[\mathsf{I}^{\mathrm{DL}}\right] = \sum_{i=1}^{K} 2\pi\lambda_i \left(\frac{\mathcal{D}_{ji}^{\frac{2-\alpha}{\alpha}} R_j^{2-\alpha}}{G(\alpha-2)}P_i + \frac{\mathcal{K}_1(d_u, \alpha_u, \Lambda_i^{\mathrm{UL}})}{G_u}\mathbb{E}\left[\Gamma_i\right]\right) \tag{20}$$

where $\mathbb{E}\left[\Gamma_i\right]$ is given in **Lemma 5** and the mean of SI power is

$$\mathbb{E}\left[\mathsf{I}_{\mathrm{SI}}^{\mathrm{DL}}\right] = \frac{\sigma_u \epsilon \alpha \rho_k G^{\epsilon}}{2(\pi\Lambda_k^{\mathrm{UL}})^{\frac{\epsilon\alpha}{2}}}\,\gamma\left[\frac{\epsilon\alpha}{2}, \pi\Lambda_k^{\mathrm{UL}}\left(\frac{P_{\max}}{\rho_k G^{\epsilon}}\right)^{\frac{2}{\epsilon\alpha}}\right]. \tag{21}$$

*Proof:* See **Appendix B-II**. ∎

The following theorem presents closed-form expressions for the mean rate utility in (13) for a generic FD link when a UE associates with the same tier $j$ with probability $\psi_{jj}$ and associates with tier $j$ for DL transmission and tier $k$ for UL transmission with probability $\psi_{jk}$.

**Theorem 1.** *(Mean rate utility of FD networks)* The mean rate utility of the FD network model described above is given by $\bar{\mathsf{R}} = \bar{\mathsf{R}}_o -$

$$\sum_{j=1}^{K}\sum_{k=1}^{K} \psi_{jk} G\left(\frac{\tau^{\mathrm{DL}}}{P_j}\left(\frac{\mathcal{A}_1(j)}{\pi\Lambda_j^{\mathrm{DL}}} + \frac{\Gamma\left[\frac{2+\alpha}{2}\right]\mathcal{A}_2}{(\pi\Lambda_j^{\mathrm{DL}})^{\frac{\alpha}{2}}} + \frac{\sigma_u\mathbb{E}[\Gamma_k]}{\mathcal{K}_3(j,k)}\right) + \frac{\tau^{\mathrm{UL}}}{\rho_k}\frac{\sigma_{b_k} P_k + \mathcal{A}_3(k)}{G^{\epsilon}(\pi\Lambda_k^{\mathrm{UL}})^{\frac{\alpha}{2}}}\mathcal{K}_4(k)\right) \tag{22}$$

where $\bar{\mathsf{R}}_o = \ln\mathsf{R}_o^{\mathrm{DL}} + \ln\mathsf{R}_o^{\mathrm{UL}}$ and

$$\mathcal{A}_1(j) = \sum_{i=1}^{K}\frac{2\pi\lambda_i\mathcal{D}_{ji}^{\frac{2-\alpha}{\alpha}}}{G(\alpha-2)}P_i, \quad \mathcal{A}_2 = \sigma^2 + \sum_{i=1}^{K}\frac{2\pi\lambda_i\mathcal{K}_1(d_u,\alpha_u,\Lambda_i^{\mathrm{UL}})}{G_u}\mathbb{E}\left[\Gamma_i\right]$$

$$\mathcal{A}_3(k) = \sigma^2 + \sum_{i=1}^{K}2\pi\lambda_i\left(\frac{\mathcal{K}_1(d_b,\alpha_b,\frac{\lambda_i}{\beta_b})}{G_b}P_i + \frac{\mathcal{U}_{ik}^{\frac{2-\alpha}{\alpha}}\mathcal{K}_2(i)}{G(\alpha-2)}\rho_i\right). \tag{23}$$

and

$$\mathcal{K}_3(j,k) = \begin{cases} \frac{1}{\Gamma\left[\frac{1+\alpha}{2}\right]}\left(\pi\sum_{i=1}^{K}\max\left\{\mathcal{D}_{ji}, \mathcal{U}_{ji}\right\}^{\frac{2}{\alpha}}\lambda_i\right)^{\frac{\alpha}{2}}, & j = k \\ \left(\sum_{l=k}^{j-1}\frac{-1}{\Omega_l}\mathcal{H}_{jl}\left[1,1;\alpha\right]\right)^{-1}\sum_{l=k}^{j-1}\mathcal{H}_{jl}\left[\mu_l, \mu_{l+1};0\right], & k < j \end{cases} \tag{24}$$

$$\mathcal{K}_4(k) = (\pi\Lambda_k^{\mathrm{UL}})^{\frac{\epsilon\alpha}{2}}\gamma\left[\frac{2+\alpha(1-\epsilon)}{2}, \pi\Lambda_k^{\mathrm{UL}}\left(\frac{P_{\max}}{\rho_k G^{\epsilon}}\right)^{\frac{2}{\epsilon\alpha}}\right] + \frac{\rho_k G^{\epsilon}}{P_{\max}}\Gamma\left[\frac{2+\alpha}{2}, \pi\Lambda_k^{\mathrm{UL}}\left(\frac{P_{\max}}{\rho_k G^{\epsilon}}\right)^{\frac{2}{\epsilon\alpha}}\right]$$

$$\overset{(P_{\max} \to \infty)}{=} \left(\pi \Lambda_k^{\mathsf{UL}}\right)^{\frac{\epsilon\alpha}{2}} \Gamma\left[\frac{2 + \alpha(1-\epsilon)}{2}\right]. \tag{25}$$

*Proof:* See **Appendix C**. ∎

Note that in (25), the first term represents the mean of the inverse of the useful signal power received at a typical BS from a UE being served that is able to perform channel inversion power control without exceeding the power budget. On the other hand, the second term represents the mean of the inverse of the useful signal power received from a UE that is transmitting with the maximum power $P_{\max}$.

*E. Special Cases*

Using the lemmas derived above and **Theorem 1**, the performance of (i) a network tier that consists only of legacy HD BSs with UL transmissions and (ii) a typical legacy HD UE with DL transmissions is presented in the following two corollaries where the proofs follow directly from (22). It is worth mentioning that if we add SI at FD BSs and FD UEs, the expressions for mean rate utility in **Corollaries 1** and **2**, respectively, will be equivalent to the mean rate utility of UL and DL transmissions in an FD network.

*Corollary 1.* (*Legacy uplink transmissions*) The mean rate utility offered by the $k$-th tier that consists only of HD BSs in an FD cellular network to its UEs is given by $\ln \mathsf{R}_k^{\mathsf{UL}} = \ln \mathsf{R}_o^{\mathsf{UL}} -$

$$\frac{\tau^{\mathsf{UL}}}{\rho_k} G^{1-\epsilon} \left(\sigma^2 + \sum_{i=1}^{K} 2\pi\lambda_i \left(\frac{\mathcal{K}_1(d_b, \alpha_b, \frac{\lambda_i}{\beta_b})\mathbf{1}\{i \neq k\}}{G_b} P_i + \frac{\mathcal{U}_{ik}^{\frac{2-\alpha}{\alpha}} \mathcal{K}_2(i)}{G(\alpha-2)}\rho_i\right)\right) \frac{\mathcal{K}_4(k)}{(\pi\Lambda_k^{\mathsf{UL}})^{\frac{\alpha}{2}}}. \tag{26}$$

*Corollary 2.* (*Legacy downlink transmissions*) The mean rate utility offered by an FD network to a legacy HD UE served by the $j$-th tier is given by: $\ln \mathsf{R}_j^{\mathsf{DL}} = \ln \mathsf{R}_o^{\mathsf{DL}} -$

$$\frac{\tau^{\mathsf{DL}}}{P_j} G \left(\frac{\Gamma\left[\frac{2+\alpha}{2}\right]}{(\pi\Lambda_j^{\mathsf{DL}})^{\frac{\alpha}{2}}}\sigma^2 + \sum_{i=1}^{K} 2\pi\lambda_i \left(\frac{\mathcal{D}_{ji}^{\frac{2-\alpha}{\alpha}}}{\pi\Lambda_j^{\mathsf{DL}}G(\alpha-2)} P_i + \frac{\Gamma\left[\frac{2+\alpha}{2}\right]\mathcal{K}_1(d_u, \alpha_u, \Lambda_i^{\mathsf{UL}})}{G_u(\pi\Lambda_j^{\mathsf{DL}})^{\frac{\alpha}{2}}}\mathbb{E}\left[\Gamma_i\right]\right)\right). \tag{27}$$

In addition, the performance of HD UL networks and HD DL networks can be obtained as presented in the following corollaries where the proofs follow directly from (22).

*Corollary 3.* (*Half-duplex uplink networks*) The mean rate utility of HD UL networks with weighted path-loss user association is given by

$$\bar{\mathsf{R}} = \ln \mathsf{R}_o^{\mathsf{UL}} - \sum_{k=1}^{K} A_k^{\mathsf{UL}} \frac{\tau^{\mathsf{UL}}}{\rho_k} G^{1-\epsilon} \left(\sigma^2 + \sum_{i=1}^{K} \frac{2\pi\lambda_i \, \mathcal{U}_{ik}^{\frac{2-\alpha}{\alpha}} \mathcal{K}_2(i)}{G(\alpha-2)}\rho_i\right) \frac{\mathcal{K}_4(k)}{(\pi\Lambda_k^{\mathsf{UL}})^{\frac{\alpha}{2}}} \tag{28}$$

$$\overset{(\sigma^2=0)}{\overset{(P_{\max}\to\infty)}{=}} \ln \mathsf{R}_o^{\mathrm{UL}} - \sum_{k=1}^{K} A_k^{\mathrm{UL}} \frac{\tau^{\mathrm{UL}}}{\rho_k} \Gamma\left[\frac{2+\alpha(1-\epsilon)}{2}\right] \Gamma\left[\frac{4-\alpha(1-\epsilon)}{2}\right] \sum_{i=1}^{K} \frac{2\rho_i\, \mathcal{U}_{ik}^{\frac{2-\alpha}{\alpha}}}{G^\epsilon(\alpha-2)} A_i^{\mathrm{UL}} \tag{29}$$

$$\overset{(\sigma^2=0)}{\overset{(P_{\max}\to\infty)}{\overset{(U_i=\rho)}{=}}} \ln \mathsf{R}_o^{\mathrm{UL}} - \frac{2\tau^{\mathrm{UL}}}{G^\epsilon(\alpha-2)}\, \Gamma\left[\frac{2+\alpha(1-\epsilon)}{2}\right] \Gamma\left[\frac{4-\alpha(1-\epsilon)}{2}\right]. \tag{30}$$

***Corollary* 4.** *(Half-duplex downlink networks)* The mean rate utility of HD DL networks with weighted path-loss user association is given by

$$\bar{\mathsf{R}} = \ln \mathsf{R}_o^{\mathrm{DL}} - \sum_{j=1}^{K} A_j^{\mathrm{DL}} \frac{\tau^{\mathrm{DL}}}{P_j} G\left(\frac{\Gamma\left[\frac{2+\alpha}{2}\right]}{(\pi\Lambda_j^{\mathrm{DL}})^{\frac{\alpha}{2}}}\sigma^2 + \sum_{i=1}^{K} \frac{2\pi\lambda_i \mathcal{D}_{ji}^{\frac{2-\alpha}{\alpha}}}{\pi\Lambda_j^{\mathrm{DL}} G(\alpha-2)}P_i\right) \tag{31}$$

$$\overset{(\sigma^2=0)}{=} \ln \mathsf{R}_o^{\mathrm{DL}} - \sum_{j=1}^{K} A_j^{\mathrm{DL}} \frac{\tau^{\mathrm{DL}}}{P_j} \sum_{i=1}^{K} \frac{2P_i\, \mathcal{D}_{ij}}{\alpha-2} A_i^{\mathrm{DL}} \tag{32}$$

$$\overset{(\sigma^2=0)}{\overset{(D_i=P_i^{-1})}{=}} \ln \mathsf{R}_o^{\mathrm{DL}} - \frac{2\tau^{\mathrm{DL}}}{\alpha-2}. \tag{33}$$

The results presented in (29) and (32) are consistent with the previous results in [7] on user association in multi-tier HD cellular networks. Furthermore, from (30) and (33), it can be seen that the performance of an interference-limited network is independent of the spatial density, receiver sensitivity, and transmit power of BSs, which is consistent with the results in existing literature [6], [7], [24], [25], [27].

In addition, by comparing the results in **Corollaries 1** and **2** with those in **Corollaries 3** and **4**, respectively, it can be seen that the mean rates of both UL and DL transmissions in FD networks are lower than those in HD networks. This degradation is due to the additional interference resulting from FD transmissions and SI. It can be seen that for UL transmissions, DL-to-UL interference can be reduced by increasing the isolation of BSs. This can achieved by increasing SIC capability of BSs or using directional antennas where both vertical and horizontal patterns are designed such that the BSs are not in the main-lobe of each other. For DL, the effect of UL-to-DL interference can be seen to be less severe because of the low transmit power of UEs compared to that of BSs.

Furthermore, the performance of FD networks with CUA and FD networks with 3NT are presented in the following two corollaries where the proofs follow directly from (22).

***Corollary 5.*** *(Coupled user association)* The mean rate utility of FD networks with CUA is given by: $\bar{\mathsf{R}} = \bar{\mathsf{R}}_o -$

$$\sum_{k=1}^{K} A_k^{\mathrm{DL}} \left( \frac{\tau^{\mathrm{DL}}}{P_k} G \left( \frac{\mathcal{A}_1(k)}{\pi \Lambda_k^{\mathrm{DL}}} + \Gamma \left[ \frac{2+\alpha}{2} \right] \frac{\sigma_u \mathbb{E}[\Gamma_k] + \mathcal{A}_2}{(\pi \Lambda_k^{\mathrm{DL}})^{\frac{\alpha}{2}}} \right) + \frac{\tau^{\mathrm{UL}}}{\rho_k} G^{1-\epsilon} \frac{\sigma_{b_k} P_k + \mathcal{A}_3(k)}{(\pi \Lambda_k^{\mathrm{DL}})^{\frac{\alpha}{2}}} \mathcal{K}_4(k) \right). \quad (34)$$

***Corollary 6.*** *(3-node topology)* The mean rate utility of FD networks with 3NT for a UE served by $j$-th tier in DL in given by (27) where that for a UE served by $k$-th tier in UL is given by $\ln \mathsf{R}_k^{\mathrm{UL}} = \ln \mathsf{R}_o^{\mathrm{UL}} -$

$$\frac{\tau^{\mathrm{UL}}}{\rho_k} G^{1-\epsilon} \left( \sigma^2 + \sigma_{b_k} P_k + \sum_{i=1}^{K} 2\pi \lambda_i \left( \frac{\mathcal{K}_1(d_b, \alpha_b, \frac{\lambda_i}{\beta_b})}{G_b} P_i + \frac{\mathcal{U}_{ik}^{\frac{2-\alpha}{\alpha}} \mathcal{K}_2(i)}{G(\alpha-2)} \rho_i \right) \right) \frac{\mathcal{K}_4(k)}{(\pi \Lambda_k^{\mathrm{UL}})^{\frac{\alpha}{2}}}. \quad (35)$$

*Proof:* The proof follows from **Corollaries 1** and **2** while considering SI affecting UL transmissions at BSs. ∎

## V. Rate Maximization in FD Networks by Optimal User Association

In this section, for optimal user association, we derive closed-form expressions for the weighting factors that maximize the mean rate utility of FD cellular networks given in **Theorem 1**. For analytical tractability, we assume that (i) all BSs have the same receiver sensitivity (i.e., $\rho_k = \rho$), (ii) there is no constraint on the maximum transmit power of UEs (i.e., $P_{\max} \to \infty$), and (iii) SI at BSs is modeled as noise such that $\sigma_{b_k} P_k = \sigma_b$. The last assumption implies that the SIC capability of a BS is proportional to its transmit power (i.e., $\sigma_{b_k} \propto \frac{1}{P_k}$), that is, the cancellation capability of BSs with higher transmit power such as macrocell BSs is better than that of BSs with lower transmit power such as picocell or femtocell BSs. Following a similar procedure as in [7], we first derive the optimal effective spatial density $\Lambda_k^{*\mathrm{UL}}$ and $\Lambda_j^{*\mathrm{DL}}$, then we retrieve the optimal weighting factors $U_k^* \leftarrow a_1 (\Lambda_k^{*\mathrm{UL}})^{\frac{\alpha}{2}}$ and $D_j^* \leftarrow a_2 (\Lambda_j^{*\mathrm{DL}})^{\frac{\alpha}{2}}$ which satisfy $\sum_{j=1}^{K} \frac{\lambda_j}{\Lambda_j^{*\mathrm{DL}}} = 1$ and $\sum_{k=1}^{K} \frac{\lambda_k}{\Lambda_k^{*\mathrm{UL}}} = 1$ for any positive constants $a_1$ and $a_2$ (cf. **Lemma 2**).

After some mathematical manipulations, maximizing (22) is equivalent to solving the following set of concave optimization sub-problem simultaneously:

$$\text{P1:} \quad \min_{\Lambda_j^{\text{DL}}, \Lambda_j^{\text{UL}}} \sum_{j=1}^{K} \frac{c_1 \lambda_j}{P_j (\Lambda_j^{\text{DL}})^{\frac{2+\alpha}{2}}} \sum_{i=1}^{K} \frac{P_i \lambda_i}{(\Lambda_i^{\text{DL}})^{\frac{2-\alpha}{2}}} + \sum_{j=1}^{K} \frac{c_2 \lambda_j}{P_j (\Lambda_j^{\text{DL}})^{\frac{2+\alpha}{2}}} \sum_{i=1}^{K} \frac{\rho_i \lambda_i \mathcal{K}_1(d_u, \alpha_u, \Lambda_i^{\text{UL}})}{(\Lambda_i^{\text{UL}})^{\frac{\epsilon\alpha}{2}}}$$

$$\text{P2:} \quad \min_{\Lambda_j^{\text{DL}}, \Lambda_j^{\text{UL}}} \sum_{j=1}^{K} \frac{c_3 \lambda_j}{\rho_j (\Lambda_j^{\text{UL}})^{\frac{2+\alpha}{2}}} \sum_{i=1}^{K} \rho_i \lambda_i \left(\Lambda_i^{\text{UL}}\right)^{\frac{\epsilon\alpha}{2}} + \sum_{j=1}^{K} \frac{c_4 \lambda_j}{P_j (\Lambda_j^{\text{DL}})^{\frac{2+\alpha}{2}}} \sum_{i=1}^{K} \frac{\rho_i \lambda_i}{(\Lambda_i^{\text{UL}})^{\frac{2+\epsilon\alpha}{2}}}$$

subject to
$$\sum_{j=1}^{K} \frac{\lambda_j}{\Lambda_j^{\text{DL}}} = 1, \quad \sum_{j=1}^{K} \frac{\lambda_j}{\Lambda_j^{\text{UL}}} = 1, \quad \text{and} \quad \Lambda_j^{\text{DL}}, \Lambda_j^{\text{UL}} \geq 0, \quad \forall j$$

where $c_i$ is an arbitrary positive constant for $i \in \{1, 2, 3, 4\}$.

Using Lagrangian relaxation and taking the first order partial derivatives with respect to $\Lambda_j^{\text{DL}}$ and $\Lambda_j^{\text{UL}}$, the optimal user association probability can be obtained as follows:

$$\Lambda_j^{*\text{DL}} = P_j^{-\frac{2}{\alpha}} \sum_{i=1}^{K} P_i^{\frac{2}{\alpha}} \lambda_i \quad \text{and} \quad \Lambda_j^{*\text{UL}} = \sum_{i=1}^{K} \lambda_i \tag{36}$$

which are equivalent to $D_j^* = \frac{D}{P_j}$ and $U_j^* = U$ for arbitrary positive constants $D$ and $U$. This result is presented in the following theorem.

**Theorem 2.** *(Optimal user association in multi-tier FD networks)* For an FD network, the mean rate utility is maximized when user association is based on maximum received power in DL and on minimum distance in UL.

It is worth mentioning that the result in **Theorem 2** shows the importance of DUA in FD cellular networks in order to maximize the overall mean rate of the network. The reason that a UE prefers to be served based on the maximum received SINR for DL and the minimum distance for UL can be explained as follows. From the DL perspective: (i) the received signal power at the UE from its serving BS is maximized, (ii) the received interference power at the UE from other BSs is minimized, and (iii) the SI power at the UE is minimized when transmitting to the nearest BS in UL. From the UL perspective: (i) the received signal power from the UE at its serving BS is maximized, (ii) the received interference power from other UEs at the serving BS is minimized when all UEs transmit with the minimum power, and (iii) since all BSs transmit with fixed power, the network tier of the serving BS for DL (or DL weighting factors) has no effect on the mean UL rate. The result above also shows the superiority of DUA compared to CUA in the FD network under consideration. That is, as shown in **Theorem 1**, two different association criteria are required for UL and DL user association to maximize the mean rate of FD networks. Clearly, this is possible only with DUA but not possible in general under any CUA criterion where the UEs are forced to associate with the same BS for both UL and DL. This result is also intuitive since CUA is a special case of DUA implying that the maximum performance of CUA can be always achieved by DUA.

## VI. Numerical Results and Discussions

### A. System Parameters

We use the results obtained above in closed-form to evaluate system performance in different scenarios. We consider the following cases: (i) FD networks with both CUA and DUA, (ii) FD networks with 3NT, (iii) HD DL networks (i.e., only DL transmissions), (iv) HD UL networks (i.e., only UL transmissions), (v) legacy DL transmissions (i.e., HD UEs with DL transmissions in an FD network), and (vi) legacy UL transmissions (i.e., HD BSs with UL transmissions in an FD network). For simulation and numerical evaluation, unless otherwise stated, we consider a two-tier network with spatial densities $\lambda_2 = 4\lambda_1$. The transmit powers of BSs are $\{P_1, P_2\} = \{37, 33\}$ dBm and the maximum transmit power of UEs is $P_{\max} = 23$ dBm. The path-loss exponents are $\{\alpha, \alpha_b, \alpha_u\} = \{4, 3.7, 4\}$, path-loss constants are $\{G, G_b, G_u\} = \{0, 30, 0\}$ dB, and the minimum distances for the pair correlation functions are $\{d_o, d_b, d_u\} = \{1, 40, 1\}$ m. For UL transmissions, the power control factor is $\epsilon = 0.9$ and all BSs have the same receiver sensitivity $\rho_i = -40$ dBm. The SIC capabilities of BSs and UEs are $\{\frac{1}{\sigma_{b_k}}, \frac{1}{\sigma_u}\} = \{70, 70\}$ dB and the noise power is $\sigma = -104$ dBm. For the evaluation of mean rate utility, SINR thresholds $\tau^{\mathrm{DL}}$ and $\tau^{\mathrm{UL}}$ are set to 0, i.e., $\mathsf{R}_o^{\mathrm{DL}} = \mathsf{R}_o^{\mathrm{UL}} = \ln(2)$ nats/sec/Hz.

### B. Validation of Analytical Results

We validate the expressions derived in **Lemmas 6** and **7** for the mean interference received at a BS and a UE, respectively. These two expressions are the keys to deriving all the following results including **Theorem 1** and they include all assumptions made throughout the analysis of the mean rate utility. The curves in Fig. 2 compare the results obtained from simulations and analysis as a function of the spatial density of tier 1. For Monte Carlo simulations, the locations of the BSs simulated as independent PPPs over a $20 \times 20$ km$^2$ area with a typical UE or BS at the origin. To simulate the active UEs that are causing UL-to-UL and UL-to-DL interference, the UEs are first dropped as a PPP with high spatial density, then the UEs associate with BSs based on the defined association criteria for UL. Then, each BS randomly selects one UE for UL transmission. Hence, **Assumption 2** is not retained in the simulation. The mean interference power for various links is averaged over $10,000$ iterations. The results in Fig. 2 validate the accuracy of our approach to derive the mean interference and show that the assumptions made above have a minor effect on the accuracy of the proposed analytical model.

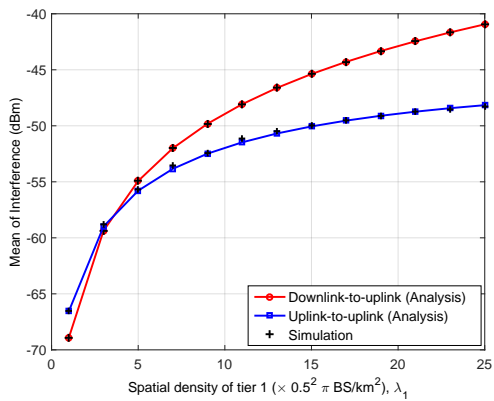

(a) Mean of Interference at a typical BS

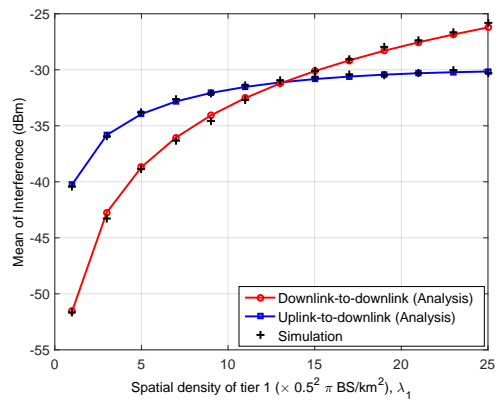

(b) Mean of Interference at a typical UE

Fig. 2. Analysis (**Lemmas 6** and **7**) vs. simulation: Mean of interference resulting from DL and UL transmissions at BSs and UEs.

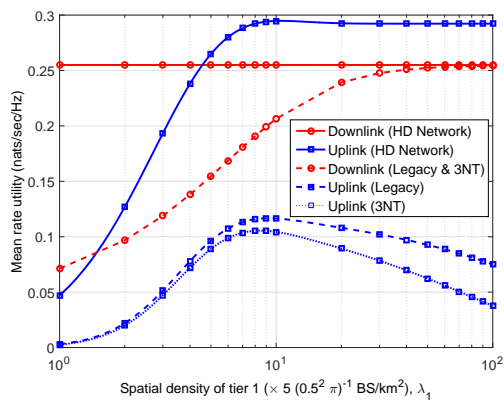

(a) Mean rate utility of DL and UL

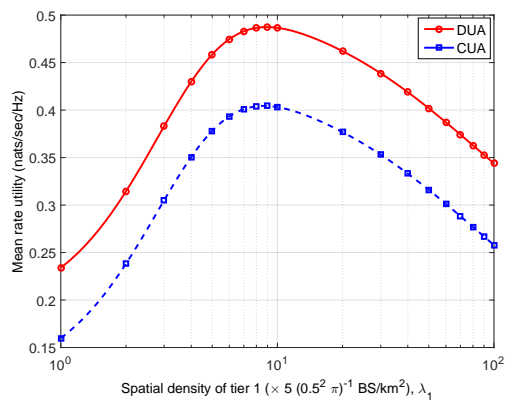

(b) Mean rate utility of DUA and CUA

Fig. 3. Mean rate utility (in nats/sec/Hz) vs. spatial density of tier 1 (in BS/km$^2$).

## C. Effect of Spatial Density

Fig. 3 shows the effect of varying the density of BSs on the mean rate utility for three different networks: (i) HD DL network, (ii) HD UL network, and (iii) FD network including both legacy DL and UL transmissions, and (iv) 3NT. For HD DL networks (i.e., **Corollary 4**) in Fig. 3a, the mean rate utility of DL transmissions is *almost* independent of the density of BSs. This can be explained as increasing $\lambda$ results in increasing the power of both the useful signal and interference and the SINR remains unchanged. For HD UL networks (i.e., **Corollary 3**), the performance results can be explained as follows. In sparse networks with low density, the distances between UEs and their serving BSs become large and the UEs transmit with their maximum power with high probability. Consequently, the useful received signal power decreases and the interference power is increased at the serving BS. Increasing the BS density improves the mean rate by making UEs closer to their serving BSs which in turn increases their ability to perform channel inversion without exceeding the maximum power budget. This increases the useful signal power and decreases interference power. With a very high BS density, the network becomes interference-limited and *almost* all UEs are able to invert their channel towards the serving BS, hence, the mean rate becomes independent of the spatial density of BSs.

Fig. 3a shows that increasing the density of FD networks has two different effects on the mean rate utility of legacy transmissions. For legacy DL transmissions (i.e., **Corollary 2**), the mean rate is lower compared to HD DL networks where the difference is due to the extra interference resulting from UL transmissions sharing the same spectrum in the FD network. However, increasing the density of BSs (i) decreases the transmit power of UEs, hence, decreases UL-to-DL interference, (ii) increases the useful signal power which is counteracted by the increase in DL-to-DL interference. Overall, this improves the mean rate of legacy DL transmissions and the performance approaches that of HD DL networks in very dense networks. On the other hand, the mean rate of legacy UL BSs (i.e., **Corollary 1**) degrades with increasing spatial density of BSs. Compared to HD UL networks, the difference is due to the extra interference resulting from DL transmissions. With increasing spatial density, the BSs become closer to each other which increases the DL-to-UL interference power received at all BSs. At the same time, the useful signal power received at any BS is upper bounded by $\rho$. Therefore, the SINR becomes very low and the mean UL rate approaches zero with increasing BS density. A similar behavior is evident for 3NT as shown in Fig. 3a. The mean DL rate is identical to that of legacy DL because both rates are evaluated at an HD UE. On the other hand, the difference between the mean UL rate in 3NT with FD BS and legacy UL with HD BS is due to the SI experienced at the FD BS in 3NT.

Fig. 3b elaborates more on the effect of spatial density in FD networks by showing the mean rate utility of a generic FD link for different user association criteria (i.e., **Theorem 1** and **Corollary 5**). Overall, it can be seen that the spatial density should be adjusted to balance the trade-off between the rates of DL and UL transmissions in FD networks. As shown in Fig. 3a, in dense FD networks, UL transmissions become more susceptible to high DL-to-UL interference. On the other hand, low spatial density of BSs degrades the performance of both DL and UL transmissions, respectively, due to high UL-to-DL interference and the maximum power budget

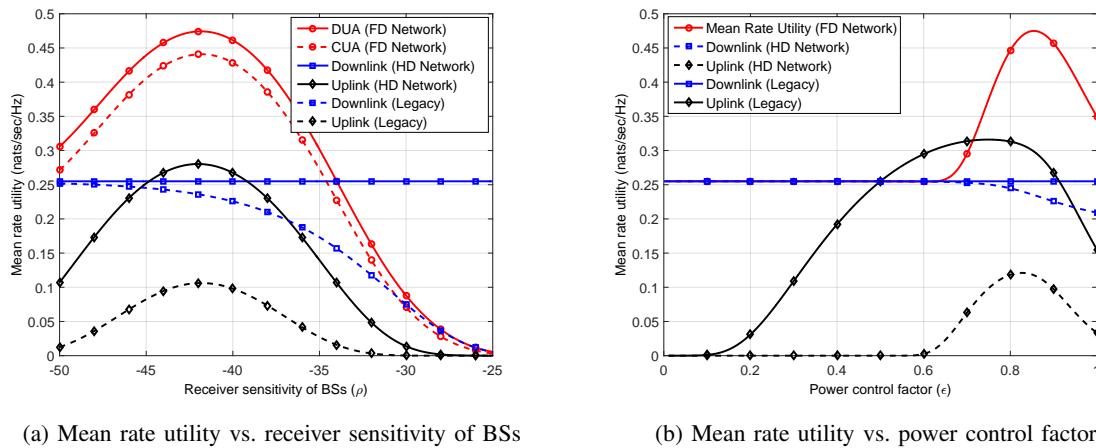

(a) Mean rate utility vs. receiver sensitivity of BSs

(b) Mean rate utility vs. power control factor

Fig. 4. Mean rate utility (in nats/sec/Hz) vs. UL power control parameters: sensitivity of BS receivers $\rho$ (in dBm) and power control factor $\epsilon$.

constraint. Therefore, as shown in Fig. 3b, as the spatial density of BSs increases, the mean rate utility of the FD network increases up to a maximum value due to the improvement in both DL and UL transmissions, then it starts to decrease due to the degradation in the mean UL rate.

### D. Effect of Power Control

Fig. 4a shows the effect of varying $\rho$ on the mean rate utility of different networks. In general, decreasing the sensitivity of the receiver (i.e., increasing $\rho$) increases the amount of transmit power required by each UE to perform channel inversion towards the serving BS (i.e., **Lemma 5**). This, in turn, increases the power of useful signal, UL-to-UL interference, and UL-to-DL interference (cf., **Lemmas 6** and **7**). Hence, as shown in Fig. 4a, the mean rate of DL transmissions in the FD network deteriorates with increasing $\rho$ because of UL-to-DL interference compared to the HD DL scenario. For UL transmissions in both HD UL and FD networks, the mean UL transmission rate increases due to increased useful signal power. This happens up to a maximum value, then the rate starts to decrease because the transmit power of UEs becomes limited by the power budget $P_{\max}$. Overall, there exists an optimal value of $\rho$ that maximizes the mean rate of FD networks and splits the performance into two regimes. That is, as $\rho$ increases, the mean rate of FD networks increases up to a maximum value, then it starts to decrease. This behavior can be explained as follows. When $\rho$ is very low (the left side of the optimal point), while DL transmissions experience low interference from UL transmissions, the interference at the BSs is very high compared to the useful signal power and thus the mean UL transmission rate is almost 0. As $\rho$ increases, the mean DL transmission rate of DL transmissions starts to

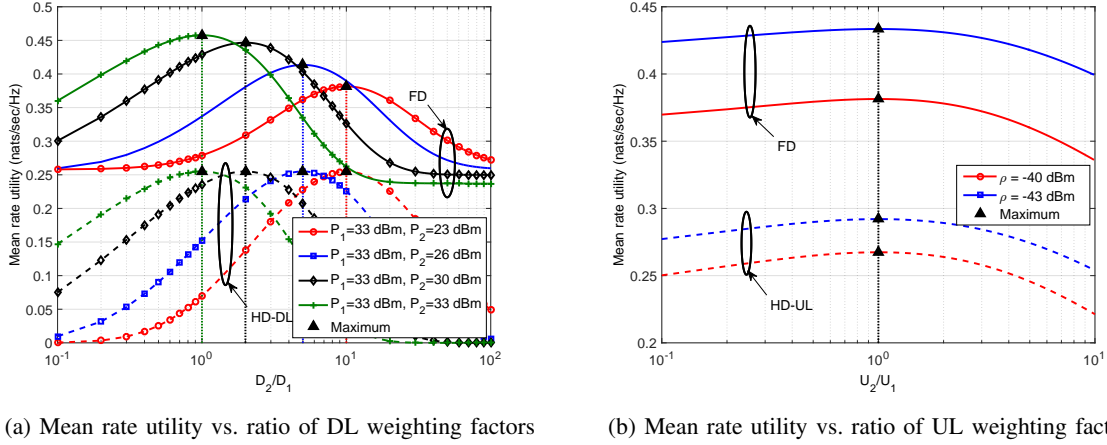

(a) Mean rate utility vs. ratio of DL weighting factors

(b) Mean rate utility vs. ratio of UL weighting factors

Fig. 5. Mean rate utility of FD, HD DL, and HD UL networks (in nats/sec/Hz) vs. the ratio of DL and UL weighting factors.

degrade while the mean rate of UL transmissions improves where this improvement dominates the overall mean rate performance. This happens until achieving the maximum mean rate. After this point, as $\rho$ increases (the right side of the optimal point), the mean rates of both DL and UL transmissions start to fall. Hence, the overall mean rate of the FD network start to decrease. A similar behavior is observed for varying the power control factor $\epsilon$ which can be justified using the same line of arguments as that for varying $\rho$. It is worth mentioning that Figs. 3b and 4a show that DUA is always superior to CUA for all ranges of $\lambda$ and $\rho$, which is in agreement with the analytical results in Section V.

### E. Effect of Weighting Factors

Fig. 5a shows the mean rate utility of FD and HD DL networks as a function of DL weighting factors for different transmit power settings. For both networks, it can be clearly seen that the maximum mean rate utility is achieved when the ratio of the weighting factors is equal to the inverse of the transmit power ratio. For example, when $\frac{P_2}{P_1} = 0.2$ (i.e., $\{P_1, P_2\} = \{33, 26\}$ dBm), the rate of FD transmissions and HD DL transmissions are maximized when $\frac{D_2}{D_1} = \frac{P_1}{P_2} = 5$. Similar remark can be made for different cases when the ratio $\frac{P_2}{P_1}$ is 1, 0.5, and 0.1. For these values of $\frac{P_2}{P_1}$, the rate is maximized when $\frac{D_2}{D_1}$ is set to 1, 2, and 10, respectively. This behavior is consistent with **Theorem 2** and can be explained as follows. For HD DL networks, intuitively, the power of the useful signal at the UE is maximized while the power of interference from other BSs is minimized when $D_j^{*\text{HD}} = P_j^{-1}$. Consequently, both the SINR and mean rate are maximized. For FD networks, same argument holds for DL transmissions as shown in **Corollary 2**. On the other hand, the mean UL transmission rate in an FD network is independent of DL weighting factors as shown in **Theorem 1** and **Corollary 1**. Hence, the mean rate utility of FD transmissions is maximized by maximizing the rate of DL transmissions by setting $D_j^{*\text{FD}} = P_j^{-1}$.

On the other hand, Fig. 5b shows the mean rate utility of FD and HD UL networks as a function of UL weighting factors for different receiver sensitivity settings. In contrast to Fig. 5a, the maximum rate utility for both FD and HD networks is achieved when the ratio of the weighting factors is equal 1. This behavior is evident for different values of $\rho$ and can be explained as follows. For HD UL networks, when the UEs associate with their nearest BSs, the aggregate interference power is minimized as the UEs transmit with the minimum power required to perform channel inversion power control. In addition, the useful signal power is maximized as the probability of a UE to perform channel inversion without exceeding $P_{\max}$ increases as the distance to the serving BS decreases. Therefore, both SINR and mean rate utility are maximum when $U_k^{*\text{HD}} = U$ for some constant $U$. For FD networks, same argument holds for UL transmissions as given in **Corollary 1**. For DL transmissions in an FD network, UL-to-DL interference is minimized when all the UEs transmit with the minimum transmit power as given in **Theorem 1** and **Corollary 2**. Hence, the mean rate utility of FD transmissions is maximized by simultaneously maximizing the rate of UL transmissions and minimizing the UL-to-DL interference where both are achieved by setting $U_k^{*\text{FD}} = U$. Therefore, it is clear that the maximum rate offered by an FD network is achieved for $\frac{D_j}{D_k} = \frac{P_k}{P_j}$ and $\frac{U_j}{U_k} = 1$. In other words, thanks to DUA, the mean rate utility can be maximized by simultaneously optimizing both DL and UL weighting factors which is not generally possible with CUA.

### F. Effect of Imperfect Self-Interference Cancellation

Fig. 6a shows the effect of SI on the performance of FD networks. It shows that the mean rate of DL and UL transmissions in FD networks are highly impacted with increasing $\sigma_u$ and $\sigma_b$, respectively. It is clear that the effect of SI is more severe for UL transmissions as it occurs at the FD BS where the transmit power is generally high compared to the power of the signal received from UL UE. Fig. 6a also shows that, based on SIC capability of UEs and BSs, HD transmissions may be preferable to FD transmissions. Fig. 6b shows the minimum required SIC capabilities of the UEs and BSs so that the rate offered by an FD network is higher than that of its HD counterpart. For example, for the case, $\lambda_2 = 4\lambda_1$, the rate offered by an HD network is higher than that offered by an FD network when the SIC capabilities of UEs and BSs are less than $40$ and $50$ dBm, respectively. In addition, SIC requirements are lower for higher spatial

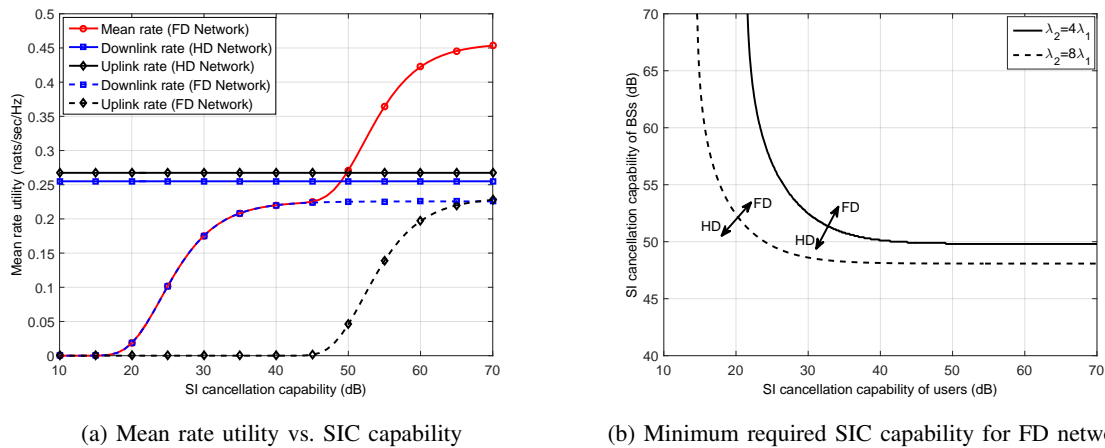

(a) Mean rate utility vs. SIC capability

(b) Minimum required SIC capability for FD networks

Fig. 6. Mean rate utility (in nats/sec/Hz) vs. SIC capability of BSs and UEs $\frac{1}{\sigma_{b_k}}$ and $\frac{1}{\sigma_u}$.

density of BSs (e.g., for $\lambda_2 = 8\lambda_1$) due to the decrease of the transmit power and consequently SI at UEs. From Fig. 6b, it can also be seen that the SIC capability of UEs does not have to be as high as that of BSs to achieve the same performance. This is mainly because the transmit power of FD UEs is not very high compared to the transmit power of BSs.

## VII. CONCLUSION

We have presented a comprehensive framework for user association in multi-tier FD cellular networks. For both UL and DL transmissions, we have considered different user association criteria including both CUA and DUA. We have used stochastic geometry to model, analyze, and evaluate the performance of the proposed system in terms of mean rate utility of FD, UL, and DL transmissions. Using weighted path-loss user association, we have derived the optimal weighting factors that maximize the mean rate utility of FD transmissions in the presence of DL-to-DL, DL-to-UL, UL-to-UL, and UL-to-DL interferences. In addition, we have shown that, in order to maximize the mean rate utility of FD networks, the UEs should associate with their nearest BSs in UL and to the BSs that result in the maximum received power in DL. This shows the advantage of using DUA over CUA. We have also shown that FD networks may be preferable to HD networks based on the level of SIC at UEs and BSs.

## Appendix A

## Proof of Association Probability and Distance Distribution

### I. Proof of *Lemma 1*

We first consider the case when a UE associates with different BSs for DL and UL. A UE associates with different BSs ($x^{\text{DL}} \in \boldsymbol{\Phi}_j$ for DL and $x^{\text{UL}} \in \boldsymbol{\Phi}_k$ for UL where $j \neq k$) under the following four conditions:

(i) $x^{\text{UL}}$ meets the criterion in (1) for UL, i.e., $U_k \|x^{\text{UL}}\|^\alpha < \min_{x \in \boldsymbol{\Phi}_i} U_i \|x\|^\alpha \quad \forall i \neq j, k$, (ii) $x^{\text{DL}}$ meets the criterion in (2) for DL, i.e., $D_j \|x^{\text{DL}}\|^\alpha < \min_{x \in \boldsymbol{\Phi}_i} D_i \|x\|^\alpha \quad \forall i \neq j, k$, (iii) $x^{\text{DL}}$ does not meet the criterion defined for UL in (1), i.e., $U_j \|x^{\text{DL}}\|^\alpha > U_k \|x^{\text{UL}}\|^\alpha$, and (iv) $x^{\text{UL}}$ does not meet the criterion defined for DL in (2), i.e., $D_k \|x^{\text{UL}}\|^\alpha > D_j \|x^{\text{DL}}\|^\alpha$.

Let $R_j = \|x^{\text{DL}}\|$ and $R_k = \|x^{\text{UL}}\|$, the event that $j \neq k$ can be expressed as:

$$\bigcap_{i=1, i \neq j,k}^{K} \left\{ \min_{x \in \boldsymbol{\Phi}_i} \|x\| > \max\left\{ \mathcal{D}_{ji} R_j^\alpha, \mathcal{U}_{ki} R_k^\alpha \right\}^{\frac{1}{\alpha}} \middle| \mathcal{D}_{jk}^{\frac{1}{\alpha}} R_j < R_k < \mathcal{U}_{jk}^{\frac{1}{\alpha}} R_j \right\}. \tag{37}$$

Hence,

$$\begin{aligned}
\psi_{jk} &\overset{(a)}{=} \mathbb{E}\left[ \exp\left[ -\pi \sum_{i=1, i \neq j,k}^{K} \max\left\{ \mathcal{D}_{ji} R_j^\alpha, \mathcal{U}_{ki} R_k^\alpha \right\}^{\frac{2}{\alpha}} \lambda_i \right] \middle| \mathcal{D}_{jk}^{\frac{1}{\alpha}} R_j < R_k < \mathcal{U}_{jk}^{\frac{1}{\alpha}} R_j \right] \\
&\overset{(b)}{=} 4\pi^2 \lambda_j \lambda_k \int_0^\infty \int_{\mathcal{D}_{jk}^{\frac{1}{\alpha}} u}^{\mathcal{U}_{jk}^{\frac{1}{\alpha}} u} uv \exp\left[ -\pi \sum_{i=1}^{K} \max\left\{ \mathcal{D}_{ji} u^\alpha, \mathcal{U}_{ki} v^\alpha \right\}^{\frac{2}{\alpha}} \lambda_i \right] \mathrm{d}v \mathrm{d}u \\
&= 2\lambda_j \lambda_k \int_{\mathcal{D}_{jk}^{\frac{1}{\alpha}}}^{\mathcal{U}_{jk}^{\frac{1}{\alpha}}} x \left( \sum_{i=1}^{K} \lambda_i \max\left( \mathcal{D}_{ji}^{\frac{2}{\alpha}}, \mathcal{U}_{ki}^{\frac{2}{\alpha}} x^2 \right) \right)^{-2} \mathrm{d}x
\end{aligned} \tag{38}$$

where $(a)$ follows from: (i) the fact that the minimum distance to a BS from a PPP $\boldsymbol{\Phi}_i$ is Rayleigh-distributed with CDF $\mathbb{P}\left[ \min_{x \in \boldsymbol{\Phi}_i} \|x\| \leq t \right] = 1 - \exp[-\pi \lambda_i t^2]$ and (ii) independence assumption for network tiers, $(b)$ follows since the expectation in $(a)$ is with respect to $R_j$ and $R_k$ which denote the distances to the serving BSs. Note also that $\psi_{jk} = 0$ when $k > j$ due to ordering the network tiers such that $\mu_i < \mu_{i+1}$ (i.e., **Assumption 1**).

For the case when $j = k$, using (1) and (2) and following a similar procedure, the probability of the event, where $x^{\text{DL}} = x^{\text{UL}} = x_o$ given that $x_o$ belongs to the $j$-th tier, can be expressed as:

$$\psi_{jj} = \mathbb{E}\left[ \exp\left[ -\pi \sum_{i=1, i \neq j}^{K} \max\{\mathcal{D}_{ji}, \mathcal{U}_{ji}\}^{\frac{2}{\alpha}} \lambda_i R_j^2 \right] \right]. \tag{39}$$

Hence, the result in (3) can be easily verified.

## II. Proof of *Lemma 4*

Following the proof in **Appendix A-I**, the joint CDF of the distance can be obtained by adding two more conditions such that $R_j > r_j$ and $R_k > r_k$. Then, the joint PDF can be obtained by differentiation. $\mathbb{E}[R_j^m]$ and $\mathbb{E}[R_k^n]$ follow the definition of the expected value.

# Appendix B

## Proof of Mean Interference

### I. Proof of *Lemma 6*

Following the definition in (10), $g_2(r)$, and $g_3(r)$, we have

$$\mathbb{E}\left[\mathsf{I}^{\mathrm{UL}}\right] \stackrel{(a)}{=} \sum_{i=1}^{K} 2\pi\lambda_i \left( \frac{P_i}{G_b} \int_{d_b}^{\infty} \frac{1 - \exp\left[-\pi\Lambda_i^{\mathrm{DL}} r^2\right]}{r^{\alpha_b - 1}}\mathrm{d}r + \mathbb{E}_{R_i}\left[ \frac{\min\{\rho_i G^\epsilon R_i^{\epsilon\alpha}, P_{\max}\}}{G} \int_{\mathcal{U}_{ik}^{\frac{1}{\alpha}} R_i}^{\infty} r^{1-\alpha}\mathrm{d}r \right] \right)$$

$$= \sum_{i=1}^{K} 2\pi\lambda_i \left( \frac{\mathcal{K}_1(d_b, \alpha_b, \Lambda_i^{\mathrm{DL}})}{G_b} P_i + \frac{\mathcal{U}_{ik}^{\frac{2-\alpha}{\alpha}}}{G(\alpha - 2)} \mathbb{E}_{R_i}\left[\min\left\{\rho_i R_i^{\epsilon\alpha}, P_{\max}\right\} R_i^{2-\alpha}\right] \right) \tag{40}$$

where $(a)$ follows from (i) the Rayleigh fading assumption of interference channels gain and (ii) Campbell's Theorem [5] knowing that the distance $R$ from the tagged BS to the closest interfering UE from tier $i$ is greater than $\mathcal{U}_{ik}^{\frac{1}{\alpha}} R_i$ (i.e., $U_i R_i^\alpha < U_k R^\alpha$) as explained earlier in Section IV-B. In addition, the spatial density of interfering BSs is $\lambda_i(1 - \exp[-\pi\frac{\lambda_i}{A_i^{\mathrm{DL}}} r^2])$. Following **Lemma 3**, we obtain $\mathbb{E}_{R_i}\left[\min\left\{\rho_i G^\epsilon R_i^{\epsilon\alpha}, P_{\max}\right\} R_i^{2-\alpha}\right] = \rho_i \mathcal{K}_2(i)$.

### II. Proof of *Lemma 7*

Following the definition in (9), $g_1(r)$, and $g_4(r)$, we have

$$\mathbb{E}\left[\mathsf{I}^{\mathrm{DL}}\right] \stackrel{(a)}{=} \sum_{i=1}^{K} 2\pi\lambda_i \left( \frac{P_i}{G} \int_{\mathcal{D}_{ji}^{\frac{1}{\alpha}} R_j}^{\infty} r^{1-\alpha}\mathrm{d}r + \frac{\mathbb{E}_{R_i}\left[\Gamma_i\right]}{G_u} \int_{d_u}^{\infty} \frac{1 - \exp\left[-\pi\Lambda_i^{\mathrm{UL}} r^2\right]}{r^{\alpha_u - 1}}\mathrm{d}r \right) \tag{41}$$

where $(a)$ follows from (i) the Rayaleigh fading assumption of interference channels gain and (ii) the fact that $D_i R^\alpha > D_j R_j^\alpha$ and the spatial density of interfering UEs is $\lambda_i(1 - \exp[-\pi\frac{\lambda_i}{A_i^{\mathrm{UL}}} r^2])$. Moreover, (21) is obtained from (11) such that $\mathbb{E}\left[\mathsf{I}_{\mathrm{SI}}^{\mathrm{DL}}\right] = \sigma_u \mathbb{E}\left[\Gamma_k\right]$.

# Appendix C

## Proof of Theorem 1

By combining (13), (15), (19), **Lemma 6**, and **Lemma 7** and rearranging all terms, we have

$$\bar{\mathsf{R}} = \sum_{j=1}^{K} \sum_{k=1}^{K} \psi_{jk}\left(\mathbb{E}_{\mathbf{R}}[\ln\mathsf{R}_{jk}^{\mathrm{UL}}] + \mathbb{E}_{\mathbf{R}}[\ln\mathsf{R}_{jk}^{\mathrm{DL}}]\right) \tag{42}$$

$$= \ln \mathsf{R}_o^{\mathrm{DL}} + \ln \mathsf{R}_o^{\mathrm{UL}} - \tau^{\mathrm{UL}} G \sum_{k=1}^{K} \frac{\sigma_{b_k} P_k + \mathcal{A}_3(k)}{\rho_k G^\epsilon} \sum_{j=1}^{K} \psi_{jk} \mathbb{E}_{\mathbf{R}} \left[ \frac{R_k^\alpha}{\min\{R_k^{\epsilon\alpha}, \frac{P_{\max}}{\rho_k G^\epsilon}\}} \right]$$

$$- \tau^{\mathrm{DL}} G \sum_{j=1}^{K} \frac{\mathcal{A}_1(j)}{P_j} \sum_{k=1}^{K} \psi_{jk} \mathbb{E}_{\mathbf{R}} \left[ R_j^2 \right] - \tau^{\mathrm{DL}} G \sum_{j=1}^{K} \frac{1}{P_j} \sum_{k=1}^{K} \psi_{jk} (\sigma_u \mathbb{E}[\Gamma_k] + \mathcal{A}_2) \mathbb{E}_{\mathbf{R}} \left[ R_j^\alpha \right] \quad (43)$$

where the expectation is with respect to the distance to the serving BS(s) (i.e., DL and UL). Hence, using **Lemmas 3** and **4**, the expressions in **Theorem 1** can be verified.